# Structural, magnetic, magnetocaloric and magneto-transport properties in Ge doped Ni-Mn-Sb Heusler Alloys

Roshnee Sahoo[a], Ajaya K. Nayak[a], K. G. Suresh[a*] and A. K. Nigam[b]

[a]*Department of Physics, Indian Institute of Technology Bombay, Mumbai-400076, India*

[b]*Tata Institute of Fundamental Research, Homi Bhabha Road, Mumbai-400005, India*

**Abstract**

The effect of Ge substitution on the magnetic, magnetocaloric and transport properties of $Ni_{45}Co_5Mn_{38}Sb_{12-x}Ge_x$ (x=0-3) has been investigated. The decrease in the exchange interaction brought by Ge substitution can be seen from the reduction in the magnetization of austenite phase and the increase in the martensitic transition temperature. Large magnetocaloric effect and magnetoresistance have been observed at room temperature, making it a potential material system for various applications.

* Corresponding author (email: suresh@phy.iitb.ac.in, FAX: +91-22-25723480)

# 1. INTRODUCTION

One of the most important properties exhibited by many ferromagnetic shape memory alloys (FSMA) is the martensitic transition, which is a first order structural transition that transforms the high temperature, high magnetic austenite phase to a low temperature, low magnetic martensite phase. As a consequence of this first order magneto-structural transition, the alloys show many interesting properties such as shape memory effect [1,2], magnetocaloric effect (MCE) [3,4] and magnetoresistance (MR) [5,6] in the vicinity of the martensitic transition. Among the various alloys of this family, NiMn based alloys have shown tremendous potential for multifunctional applications. In this series, Ni-Mn-Ga [7], Ni-Mn-Sn [3] and Ni-Mn-Sb [8] are found to exhibit promising MCE. The highest MCE of -86 J/kg K is observed in $Ni_{50}Mn_{20}Ga_{25}$ single crystals in 50 kOe field [7]. In the NiMn based Heusler alloys, the magnetic properties are mainly attributed to the Mn magnetic moments as Ni atoms have small or negligible moment [9]. It was reported (Ni,Co)-Mn-Sb Heusler system, which undergoes a first order martensite transition near room temperature shows MCE value of 9.2 J/kg K in 50 kOr field [10]. A maximum MCE value of 14.1 J/kg K has been reported at low field (10 kOe) for $Ni_{44-x}Co_xMn_{45}Sn_{11}$ Heusler system [11].

The systems with first order transition draw considerable attention for achieving giant MCE. However, determination of the magnetic entropy change using the usual method of Maxwell's relation is questionable. Recently it was shown that both phase coexistence and hysteresis associated with the first order transition would overestimate the magnetic entropy change when Maxwell's relation is used [12,13]. However, for a quick search of potential MCE materials, indirect method using Maxwell's relation is still being used [14]. In such cases, generally, the MCE is calculated with the magnetization data collected in the increasing field and decreasing field modes.

As part of our investigations on similar alloys, we have recently reported the effect of doping Si and Ga in Sb site in NiCoMnSb system [15]. Large MCE value of 70 J/kg K was obtained with Si substitution, while it was much smaller with Ga. With the aim of identifying materials with large MCE near room temperature, we have doped Ge for Sb in NiCoMnSb system and the results are presented in this paper. We present the effect of Ge substitution on the magnetic, magneto-transport and magnetocaloric properties in $Ni_{45}Co_5Mn_{38}Sb_{12}$ alloys.

## EXPERIMENTAL DETAILS

Polycrystalline samples of $Ni_{45}Co_5Mn_{38}Sb_{12-x}Ge_x$ (x=0, 1, 2, 3) were prepared by arc melting process in argon atmosphere. The constituent elements were of at least 99.9% purity. The ingots were remelted several times and the weight loss after the final melting was found to be negligible. For homogenization, the as-cast samples were sealed in evacuated quartz tube and subsequently annealed for 24 hr in 850 ˚C. The structural characterization was done by taking the powder x-ray diffraction (XRD) using Cu-Kα radiation. The magnetization (M) measurements were carried out using a vibrating sample magnetometer attached to a Physical Property Measurement System (Quantum Design, PPMS-6500) and the electrical resistivity (ρ) measurements were performed by four probe method using PPMS.

## RESULTS AND DISCUSSION

The XRD patterns for $Ni_{45}Co_5Mn_{38}Sb_{12-x}Ge_x$ were taken at room temperature. In figure 1(a) it is shown that the alloy with x=0 shows austenite (cubic) phase while alloys with x=1 and 3 show some martensite (orthorhombic) phase along with the austenite phase at room temperature. This suggests that the martensite phase gradually gets stabilized with Ge substitution.

Figure 2(a) shows the field cooled cooling (FCC) and field cooled warming (FCW) magnetization curves measured in 1 kOe field. The ferromagnetic to paramagnetic transition of the austenite phase, ($T_C^A$, not shown in figure) occurs at around 340 K. When the temperature is reduced below $T_C^A$ the magnetization reaches its maximum at the martensitic start temperature ($M_S$) and then suddenly decreases to attain its minimum value at the martensitic finish temperature ($M_F$). Below $M_F$ there is another transition at $T_C^M$ at the Curie temperature of the martensite phase. The sudden decrease of magnetization around the martensitic transition temperature indicates the presence of some antiferromagnetic (AFM) component below this transition temperature [3]. The hysteresis observed between FCC and FCW curves confirms the first order nature of the transition in the system. The most remarkable feature in the system is the reduction of hysteresis with Ge substitution. It is observed that for x=0, the hysteresis between FCC and FCW curves is around 20 K, whereas at x=3, this decreases to 10 K. It is also observed that the martensitic transition temperatures $M_S$ and $M_F$ shift to higher temperatures with the substitution of Ge. Such a trend also points to the earlier observation (seen from the XRD data) that the martensite phase stabilizes with the addition of Ge. Due to the smaller ionic radius of Ge as compared to that of Sb, Ge substitution would cause a lattice contraction, which leads to the increase in martensitic transition temperature in the system. This is consistent with the variation of martensitic transition temperature with hydrostatic pressure in $Ni_{45}Co_5Mn_{38}Sb_{12}$, in which pressure stabilizes the martensitic phase as in the present case [16]. Therefore, the effect of chemical pressure brought about by Ge seems to play a similar role as that of hydrostatic pressure.

The electrical resistivity of $Ni_{45}Co_5Mn_{38}Sb_{12-x}Ge_x$ alloys with x=0, 1 and 3 measured in 0 and 50 kOe is shown in Fig. 2(b). It is observed that the martensite phase has higher resistivity than the austenite phase. The martensite to austenite phase transition is followed by a sharp drop in resistivity. This discontinuity corresponds to a transformation from a disordered martensite state to a more ordered austenite state, as observed from magnetization data. The sharp drop of resistivity may be due to presence of super zone gaps associated with the structural transition, which alters the density of electronic states near the Fermi surface. This is also seen in NiMnIn and NiMnSn alloys [17,18], where the variation of density of states also occur in the vicinity of Fermi level. It is clear from Fig. 2b that with the application of the field, the martensitic transition is found to shift to lower temperatures in all the three alloys. However, for x=3 the shift is quite small compared to that in the other two cases.

The hysteresis between the heating and the cooling curves signifies the first order nature of the martensitic transition, as observed in the M(T) and the ρ(T) curves. The resistivity in the martensite phase is temperature independent. This may be due to large disorder in the system. It can be inferred from the figure 2(b) that the disorder increases with increase in Ge concentration. The noticeable feature that can be observed for x=3 is the broadening of the martensitic transition which signifies the decrease of discontinuity near transition region. This suggests that the sharpness of martensitic transition decreases gradually with Ge substitution, which also indicates the increased disorder at higher Ge concentrations. From the figures 2a and 2b, it can be seen that both the M-T and the $\rho$-T data are in agreement with each other.

Figure 3 represents isothermal M(H) curves for x=1 and 3 taken around the martensitic transition temperature. The measurements were done with the temperature increment $\Delta T$=1 K, upto a maximum field of 50 kOe. It is observed that the magnetization increases with increase in temperature for both the alloys. The observation of metamagnetic transition in the temperature range of 270 -274 K for x=1 signifies the field induced structural transition from the low magnetic martensite phase to the high magnetic austenite phase. However, the sharpness of metamagnetic transition significantly reduces for x=3. Near the martensitic transition, the magnetization difference (at the highest field) between two consecutive isotherms decreases with Ge substitution. The hysteresis between increasing and decreasing field modes is also significantly reduced with Ge substitution.

In order to get more insight about the order of transition, Arrott plots have been plotted and are shown in Fig. 4. From the figure 4(a), it can be noted that the $M^2$ vs. H/M curve is S shaped, which implies that the alloy undergoes a first order metamagnetic transition [19]. However, for x=3 the curves are non linear but does not show similar S shaped behavior, which suggests the reduction in the strength of the metamagnetic behavior and suppression of the discontinuity in the reverse martensitic transition.

The isothermal magnetic entropy change $\Delta S_M$ calculated from M(H) curves taken around the martensitic transition is shown in Fig.5. The $\Delta S_M$ in the present case is calculated from the M-H data in the heating mode, using the Maxwell relation [3,14]. The temperature dependence of $\Delta S_M$ is shown with applied field of $\Delta H$=10, 20, 30, 40 and 50 kOe for x=0, 1 and 3. The observed maximum positive magnetic entropy change of 68 J/kg K in x=0 decrease to 39 J/kg K and 28 J/kg K for x=1 and x=3 respectively, under a field change of 50 kOe. This decrease is due to the decrease in $\partial M / \partial T$ value and a decrease in magnetization of the austenite phase as seen from Fig. 2(a). The decrease in the magneto-structural coupling brought about by the Ge substitution may also responsible for the large decrease in the MCE value. The refrigerant capacity (RC) estimated by integrating $\Delta S_M$ curve over full width half maximum for x=1 and 3 is found to be 92 and 77 J/kg respectively for a field change of 50 kOe. The average hysteresis loss at the MCE peak temperature regime is found to be 39 and 23 J/kg for x=1 and 3 respectively. This gives rise to the effective RC value of 53 J/kg for x=1 and 54 J/kg for x=3. This RC value is larger than in the reported NiMnSb systems [4, 8].

It is important to note here that calculating MCE using Maxwell relation near first order transition is still controversial, as mentioned earlier. The fact is that the hysteresis observed near martensitic transition accounts for the irreversibility of thermodynamic variables. Therefore, the calculated quantities such as the entropy change depend on the field and the thermal histories of the system. In order to find out this dependence, the magnetic entropy change for x=1 was estimated in the increasing/decreasing fields or increasing/decreasing temperatures [20-22]. The magnetic entropy change estimated in the cooling mode for x=1 is shown as the dashed curve in figure 5(b). The entropy change thus obtained is found to be a maximum at 272.5 K, with a magnitude of 42 J/kg K for 50 kOe. This $\Delta S_M$ value is approximately same as the value obtained in the heating mode and as expected, the peak position has shifted to lower temperature in the cooling mode. Therefore, the difference in magnetic entropy change as observed for the heating and the cooling modes in x=1 is quite nominal.

The variation of magnetoresistance (MR) with temperature for x=0, 1 and 3 is shown in Fig 6. The MR is calculated using the relation MR = $[\{\rho(H,T)- \rho(0,T)\}/ \rho(0,T)]\times100$. A maximum MR of 39% is obtained for x=1 around the martensitic transition. MR of 38% and 10% is calculated for x=0 and x=3 respectively. It can be seen that the MR value for x=3 is quite small compared to that of the other two concentrations, though there is no considerable decrease in the entropy change for the same. The large value of the MR observed for x=0 and 1 is due to the significant enhancement of the FM phase due to the reverse martensitic transition induced by the field. The lower MR for x=3 is may be due to the change in electronic structure in the austenite phase as compared to martensite phase due to Ge substitution, which results in change of the density of states near Fermi surface and affects the electronic transport properties.

Based on the results obtained in this system, it is clear that the most exciting feature in the system is that the substitution of nonmagnetic element Ge for Sb gives rise to significant changes in the magnetic and transport properties. It can be seen that Ge substitution is able to stabilize the martensite phase and also reduce the magnetostructural coupling in the system. To understand this trend in terms of the number of valence electrons, the (e/a) ratio is evaluated for $Ni_{45}Co_5Mn_{38}Sb_{12-x}Ge_x$ system. The values are obtained to be 8.21, 8.20 and 8.18 for x=0, 1 and 3 respectively. It is well known that the martensite transition takes place when the Fermi surface touches the Brillouin zone [23]. Decrease in the valence electron per atomic ratio leads to a reduction in the instability of the high temperature phase. Therefore, martensite transition temperature ($T_M$) should follow e/a ratio [24] and must decrease with increase in the Ge concentration. However, in the present case the $T_M$ does not follow the (e/a) ratio and shows the opposite trend. Therefore, the variation of ionic radii may play an important role in determining the martensitic transition [25]. It is known that Ge has less ionic radius than Sb, which may result in a decrease in the cell volume. This would cause a shift of $T_M$ towards higher temperature as AFM exchange strengthens with decrease in Mn-Mn distance. As mentioned earlier, application of pressure was found to enhance the $T_M$ in NiCoMnSb Heusler alloys [16]. Therefore, the chemical substitution yields the same trend as that of hydrostatic pressure in the system.

By comparing the present results with those obtained in Si/Ga doped $Ni_{46}Co_4Mn_{38}Sb_{12}$ [15], it can be seen that (e/a) ratio decreases in all the three cases, but only in Si, the $T_M$ follows the (e/a) ratio. The ionic radius may be playing a role in the case of Ga substitution. As Ga has a smaller ionic radius than Sb, this may result in a reduction in the Mn-Mn distance. Hence martensitic transition stabilizes and shifts to higher temperatures with Ga. Therefore martensitic transition does not only depend on conduction electron density or the (e/a) ratio, it also depends ionic radius of substituted atoms. However the trend seen in the case of Si doping cannot be explained with e/a ratio and the atomic radius. Hence to get a understanding on the general dependence of the martensitic transition on the substituent atoms, other parameters have to be considered and further studies needed to be carried out.

## CONCLUSIONS

In conclusion, the magnetic, transport and magnetocaloric properties of $Ni_{45}Co_5Mn_{38}Sb_{12-x}Ge_x$ alloys have been investigated. It is observed that with Ge substitution in Sb site, the martensite phase gets stabilized and the magnetization in the austenite phase reduces systematically. The hysteresis observed for x=0 between the heating and the cooling curves in the magnetization data decreases considerably for x=3. The reduction in the sharpness of the metamagnetic transition is also observed with Ge doping. From the resistivity data it is found that the disorder in the martensite phase increases with increase in the Ge concentration. It is also found that the martensitic transition can be tuned over a large temperature interval with a small change in Sb/Ge ratio, giving rise to a large MCE and a large MR value over a broad temperature interval. Therefore, the present system seems to be a promising candidate as a multifunctional system.

**Figure Captions**

Figure.1. X-ray diffraction pattern for x=0, 1 and 3 in $Ni_{45}Co_5Mn_{38}Sb_{12-x}Ge_x$ alloys.

Figure. 2. (a) Temperature dependence of magnetization in 1 kOe field. The filled symbols represent FCC and the open symbols represent FCW data (b) Temperature variation of electrical resistivity for x=0, 1 and 3 for 0 and 50 kOe in $Ni_{45}Co_5Mn_{38}Sb_{12-x}Ge_x$ alloys.

Figure. 3. Field dependence of magnetization for (a) x=1 and (b) x=3 in $Ni_{45}Co_5Mn_{38}Sb_{12-x}Ge_x$ alloys.

Figure. 4. Arrott plots for (a) $Ni_{45}Co_5Mn_{38}Sb_{12}$ and (b) $Ni_{45}Co_5Mn_{38}Sb_9Ge_3$

Figure. 5. Temperature dependence of magnetic entropy change $\Delta S_M$ for x=0, 1 and 3

Figure. 6. Temperature dependence of magnetoresistance for x=0, 1 and 3 in $Ni_{45}Co_5Mn_{38}Sb_{12-x}Ge_x$ alloys.

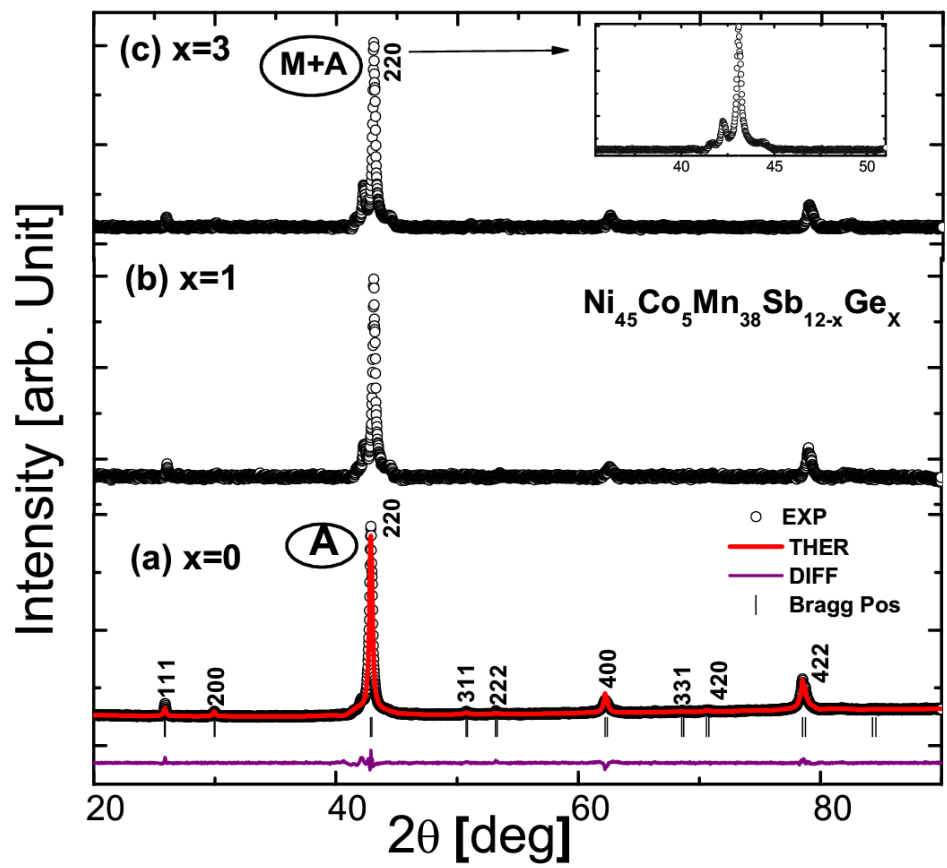

**Figure.1**

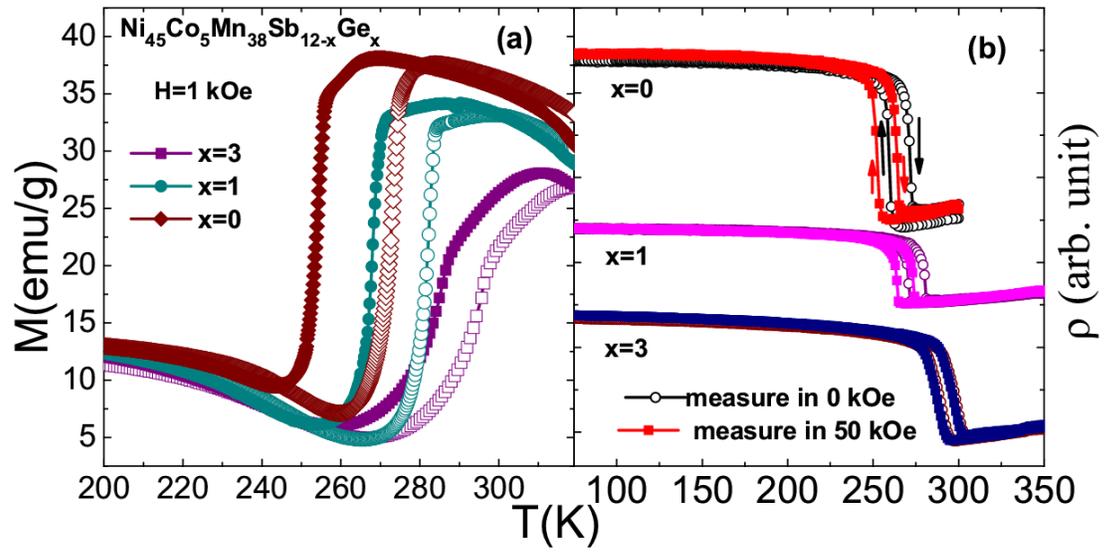

**Figure.2**

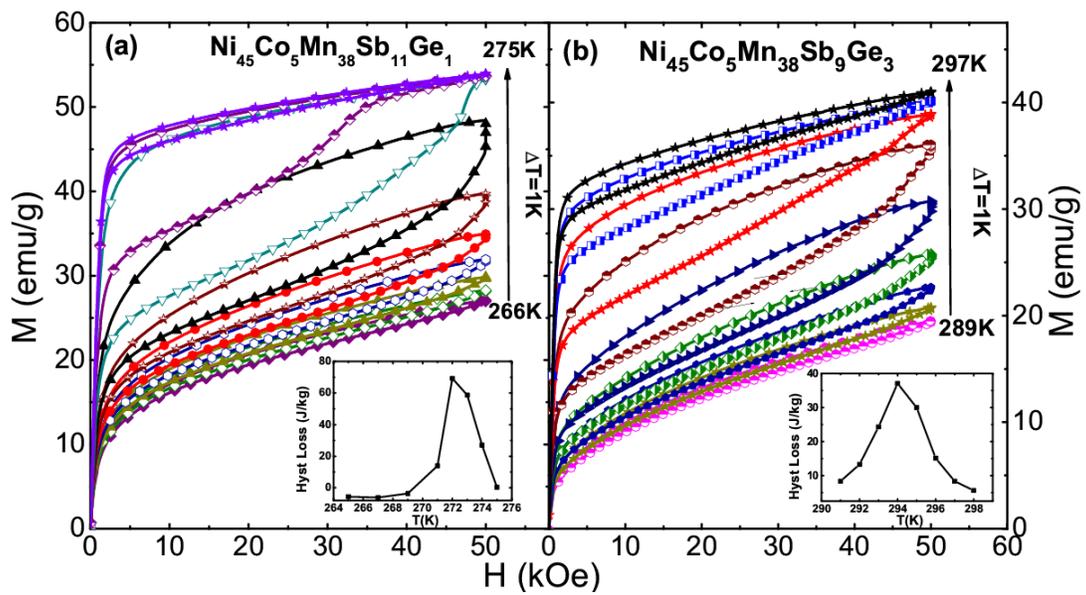

**Figure.3**

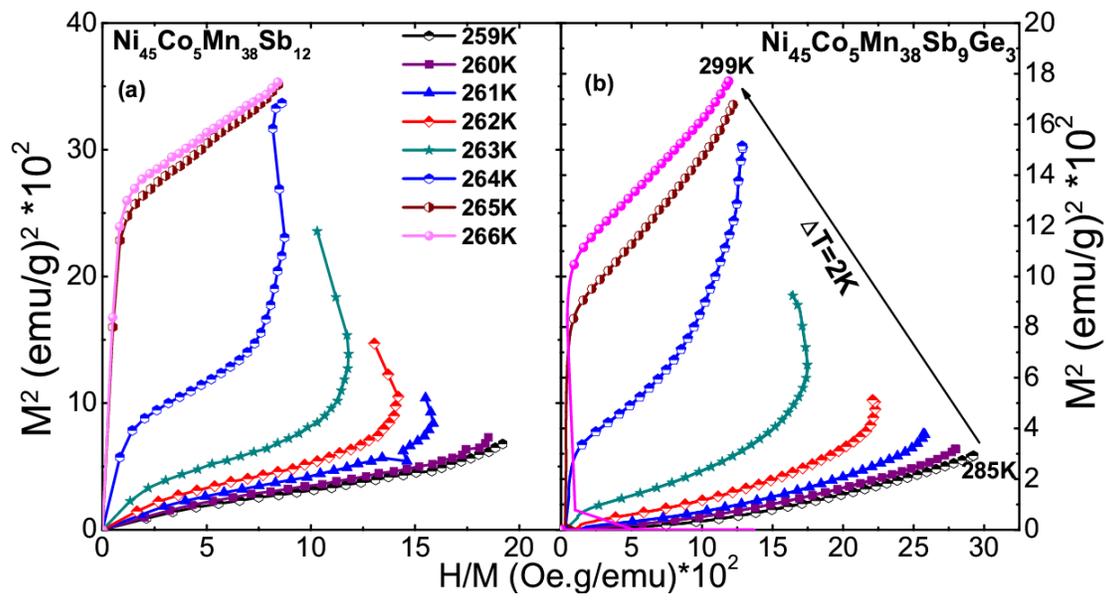

**Figure.4**

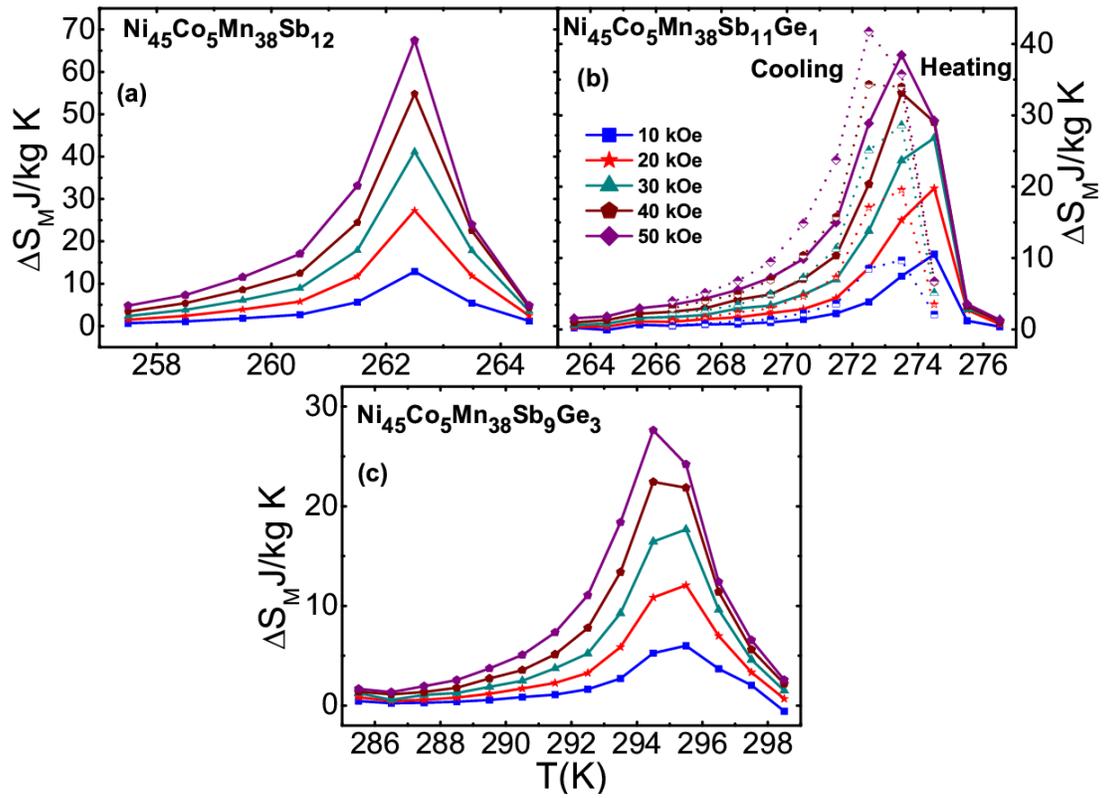

**Figure.5**

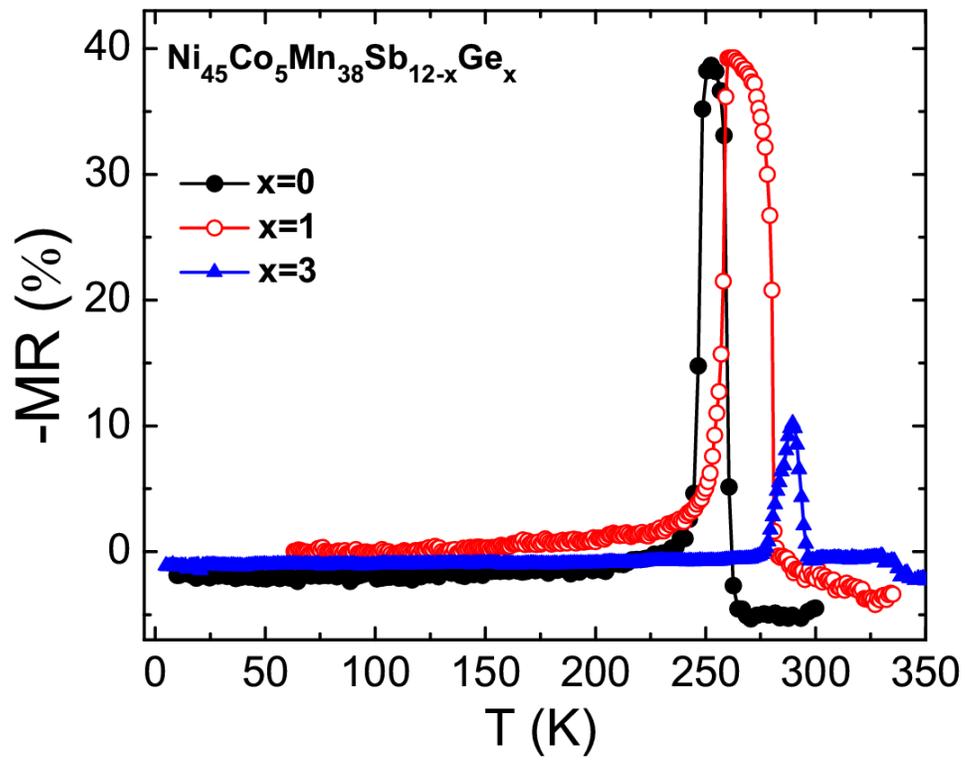

**Figure.6**